\documentclass[acmsmall]{acmart}
\usepackage{enumitem}
\usepackage{xspace}

\usepackage{xcolor}
\usepackage{array}
\usepackage{wrapfig} 
\usepackage{colortbl}

\usepackage{hyperref}

\usepackage{tabularray}
\usepackage{multirow}
\usepackage{tcolorbox}
\usepackage[normalem]{ulem}
\usepackage{pdfpages}
\definecolor{aliceblue}{rgb}{0.9, 0.97, 1.0}
\definecolor{myblue}{rgb}{1, 0, 0}

\definecolor{light-blue}{rgb}{0.6,0.6,1}
\AtBeginDocument{%
  \providecommand\BibTeX{{%
    \normalfont B\kern-0.5em{\scshape i\kern-0.25em b}\kern-0.8em\TeX}}}

\newcommand{\tool}[0]{LintCFG\xspace}

\newcommand{\zc}[1]{\textcolor{red}{\textbf{ZC:} #1}}
\newcommand{\rwr}[1]{\textcolor{red}{\textbf{Reviewer:} #1}}
\newcommand{\zj}[1]{\textcolor{blue}{\textbf{Zejun:} #1}}

\begin{document}


\title{Still Manual? Automated Linter Configuration via DSL-Based LLM Compilation of Coding Standards} 






\author{Zejun Zhang, Yixin Gan, Zhenchang Xing, Tian Zhang, Yi Li, Qinghua Lu, Xiwei Xu, Liming Zhu} 

\begin{abstract} 
Coding standards are essential for maintaining consistent and high-quality code across teams and projects. 
Linters help developers enforce these standards by detecting code violations. 
However, manual linter configuration is complex and expertise-intensive, and the diversity and evolution of programming languages, coding standards, and linters lead to repetitive and maintenance-intensive configuration work. 
To reduce manual effort, we propose \textsf{\tool}, a domain-specific language (DSL)-driven, LLM-based compilation approach to automate linter configuration generation for coding standards, independent of programming languages, coding standards, and linters. 
Inspired by compiler design, 
we first design a DSL to express coding rules in a tool-agnostic, structured, readable, and precise manner. 
Then, we build linter configurations into DSL configuration instructions. 
For a given natural language coding standard, the compilation process parses it into DSL coding standards, matches them with the DSL configuration instructions to set configuration names,  option names and values, verifies consistency between the standards and configurations, and finally generates linter-specific configurations. 
Experiments with Checkstyle for Java coding standard show that our approach achieves over 90\% precision and recall in DSL representation, with accuracy, precision, recall, and F1-scores close to 70\% (with some exceeding 70\%) in fine-grained linter configuration generation. 
Notably, our approach outperforms baselines by over 100\% in precision. 
An ablation study confirms the effectiveness of the main components of our approach. 
A user study further shows that our approach improves developers' efficiency in configuring linters for coding standards. 
Finally, we demonstrate the generality of the approach by generating ESLint configurations for JavaScript coding standards, showcasing its broad applicability across other programming languages, coding standards, and linters. 

\end{abstract}

\keywords{Coding Standard, Linter Configuration, Domain-Specific Language}

\ccsdesc[500]{Software and its engineering}



\maketitle

\section{Introduction}

Coding standards, also known as coding conventions or coding styles, are guidelines that govern how code should be written and organized within programming languages~\cite{goncharenko2016language, ogura2018bring, allamanis2014learning, smit2011code, simmons2020large, boogerd2008assessing}. 
They cover a wide range of aspects, including file basics, file structure, formatting, documentation, naming conventions, language features, and basic programming practices. 
To maintain software quality and enable long-term maintenance, many organizations, software projects, and open-source communities establish their own coding standards~\cite{gleStyle,codingguidelines,aliaba,gitlab}. 
To enforce coding standards, numerous linting tools have been developed to automatically detect standard violations in code~\cite{eslint,ckstyle,findbugs,yiu2023checkstyle,dilruk2019coding,livshits2005finding,christodorescu2003static,tsantalis2008jdeodorant,latappy2023mlinter,oumarou2015identifying,pmd}. 
Previous research has primarily focused on individual studies of coding standards and linters, such as the differences and evolution of coding standards for a programming language~\cite{abdallah2017java, campos2019mining, latappy2023mlinter, allamanis2014learning, markovtsev2019style, smit2011code}, and improving the capabilities of linting tools~\cite{beller2016analyzing, zampetti2017open, christakis2016developers, eslint, ckstyle, findbugs, pmd, tomasdottir2020adoption, tomasdottir2017and}. 

Automatically generating linter configurations for coding standards has long been overlooked, despite its significance.  
First, different programming communities, organizations, and software projects choose their own coding standards and linters. 
This creates substantial mental overhead for developers, as there is no unified tooling to streamline configuration generation. 
Secondly, the configuration process is complex: developers must first master the coding standard and linters, then sift through hundreds of configurations to find the corresponding linter configuration names, determine the matched options and their values, and then generate machine-readable formats supported by the linter. 
Moreover, developers must continuously track changes in coding standards and linter tools, which is both cumbersome and error-prone. 
As new coding standards emerge or projects transition to new linters, manually updating configurations becomes even more challenging.

Our observation on over 1,000 GitHub repositories highlights the widespread and active evolution of both coding standards and linters (see \textcolor{blue}{\href{https://github.com/anynomousaccount/lint_gen_kod/blob/main/Appendices.pdf}{APPENDIX A}}). 
From 2009 to 2024, 1,066 coding standard projects and 1,692 linter projects were created. 
Of these, 20\% (218 out of 1,066) of coding standards and 47\% (787 out of 1,692) of linter tools were under active development in 2024. 
Moreover, our analysis of Stack Overflow questions (Section~\ref{motivation}) reveals that users often do not know whether linters support a coding standard and struggle to configure linters correctly.
This highlights the need for automated linter configuration support to reduce manual efforts.

To automatically generate linter configurations to comply with a coding standard, we propose \tool, a domain-specific language (DSL)-driven, LLM-based compilation approach that is independent of programming languages, coding standards, or linters.
Our approach is inspired by compiler design~\cite{louden1997compiler,grune2012modern,muchnick1997advanced}, which focuses on translating high-level representations into machine-readable code.



%
We first design a DSL to represent coding rules in a tool-agnostic, structured, precise, and readable manner.
This decision is motivated by the observed limitations of existing representations: natural-language coding standards are tool-agnostic and readable but lack structure and precision, whereas machine-readable linter configuration formats are precise and structured but tied to specific linters and often difficult to understand. 
We use a card sorting approach~\cite{card_sorting} to design the DSL by analyzing coding rules from the Google Java coding standard and Checkstyle documentation~\cite{doc_ckstyle,gleStyle}. 
We select these sources not to serve specific standards or linters, but because of their broad adoption and comprehensive coverage, making them ideal foundations for DSL design. 
Leveraging this generic DSL and the generation capabilities of LLMs, coding rules from coding standards and linter configurations can be parsed into a unified intermediate form, agnostic to programming languages, coding standards, or linters.

We then build linter configurations as a DSL configuration instruction set. 
A typical linter configuration consists of a configuration name, option names, and their corresponding values. 
Inspired by the structure of configuration schemas, we define each configuration as a DSL instruction comprising the general instruction and the option instruction. 
The general instruction is to represent the overall behavior for configuration names in DSL. 
The option instruction is to represent the specific behavior for different option values of a option name in DSL. 
After building the linter instruction framework, we prompt LLMs to generate the DSL configuration instruction for each linter configuration based on the linter documentation. 

Finally, given a natural language (NL) coding standard, we design a compilation process to generate the linter-specific configuration building on the DSL and DSL configuration instruction set.
Similarly to compiler steps such as syntax parsing, intermediate code generation, semantic analysis, and machine code generation, we break the compilation into five steps, with each step focusing on a specific task to ensure correctness and scalability. 
First, we parse the NL coding standard into the DSL coding standard. 
Then, we select configuration names by matching the DSL coding standard with general DSL instructions.
Next, we determine option names and values by matching the option DSL instructions with the DSL coding standard. 
To ensure alignment, we verify consistency between the DSL coding standard and the matched DSL instructions in terms of rule types, checked objects, and semantics. 
Finally, we convert the aligned configurations into the linter-specific configuration (e.g., in XML format). 

 We comprehensively evaluate our approach across five dimensions. 
 We first assess the effectiveness of DSL representations for the Google Java coding standard and Checkstyle configurations. 
 The generated DSL representations achieve over 80\% accuracy and exceed 90\% in precision, recall, and F1-score. 
 Next, we evaluate the effectiveness of our approach in generating Checkstyle configurations for Google Java coding standard at three levels of granularity: configuration name, option name, and option value. 
 We design six baselines with varying levels of linter information and Retrieval-Augmented Generation (RAG) capabilities. 
 Our approach significantly outperforms all baselines, approaching or exceeding 70\% in all metrics across all levels, with minimum improvements of 51.5\%, 106.3\%, 5.1\%, and 77.7\% in accuracy, precision, recall, and F1-score, respectively. 
 Notably, precision improves by more than 100\% over the best-performing baseline. 
We further conduct an ablation study to validate the contributions of the DSL design and the compilation process. 
Furthermore, a user study shows that our generated configurations for coding standards can help linter users complete the task more quickly and correctly. 
Finally, we evaluate the generality of our approach by generating ESLint configurations for JavaScript coding standards.  
Our approach consistently achieves high accuracy, precision, recall, and F1-score, demonstrating its effectiveness across different programming languages, coding standards, and linter tools. 

The contributions of this paper are as follows:
 
\noindent  \scalebox{1.2}{\textbullet} To the best of our knowledge, we are the first to automatically generate linter configurations for coding standards.


\noindent  \scalebox{1.2}{\textbullet} We propose a DSL-driven, LLM-based compilation approach that is independent of specific programming languages, coding standards, and linter tools. 
Experimental results demonstrate the effectiveness, usefulness, and generality of the approach. 
        


\noindent  \scalebox{1.2}{\textbullet} We highlight key implications for future research and construct benchmarks for Checkstyle configurations for Google Java coding standard, as well as ESLint configurations for the Google JavaScript coding standard, to support future studies.

\vspace{-1mm}
\section{Motivation Study}\label{motivation}
\noindent \textbf{Motivation.} We are interested in exploring whether users struggle with configuring linter tools to motivate the task of automating linter configuration generation for coding standards. 

\noindent \textbf{Method.} \label{motiv_challge_so_appraoch} 
To understand the challenges of configuring linters for coding standards, we search Stack Overflow questions using linter tool names (e.g., Checkstyle, ESLint, Pylint, and Prettier) and coding standards (e.g., coding standard, coding convention, style guide and rule) as keywords. 
From the top 500 questions returned, two authors independently review and classify each question as related or unrelated to challenges in configuring linters for coding standards. 
Cohen’s kappa~\cite{viera2005understanding} is 0.89, indicating substantial agreement. 
Disagreements are resolved through discussion, resulting in 90 questions being classified as related to challenges in configuring linters for further analysis. 
Finally, the authors categorize these questions into different types of challenge through discussion. 

\noindent \textbf{Result.} We summarize two types of challenges in configuring linters for coding standards. 

\noindent \textbf{(1) Users feel uncertain whether a linter supports a coding standard} because they must fully understand each configuration's semantics and continuously track tool updates. 
It accounts for 44.4\% (40 of 90 questions). 
For example, users want Checkstyle to prohibit all class methods except specific ones~\cite{so_ckstyle_1}. 
They found a possible configuration, but it cannot exclude certain methods, making them uncertain about Checkstyle's support for the standard. 
Although this question~\cite{so_ckstyle_1} was asked 14 years ago, it was still active within 1 year and was viewed more than 291,000 times,  indicating that users find it challenging and time-consuming to determine whether linting tools support the intended coding standards. 
For another example, users asked whether Pylint supported comment spell checking 11 years ago~\cite{so_pylint_1}. 
This feature was initially unavailable but was later introduced in version 1.4, showing that users must track tool updates to determine if a linter supports a coding standard. 

\noindent \textbf{(2) Users struggle to configure the linter correctly} due to the challenges of setting the correct tool configuration names, option names, and option values for the coding standards, accounting for 55.6\% (50 of 90 questions). 
For example, users asked how to remove semicolons in Prettier. 
Although they knew this feature was supported in Prettier, they didn't know how to enable it in the settings~\cite{so_prettir_1}. 
The corresponding configuration name for Prettier is ``semi''. 
For another example, users mistakenly thought the ``exception'' option of spaced-comment rule in ESLint addressed exceptions for spaced comments on special comments~\cite{so_eslint_1}. 
In reality, it is the ``markers'' option that handles this coding standard. 

\vspace{-2mm}
\begin{tcolorbox}[width=5.5in,boxsep=2pt,left=1pt,right=1pt,top=1pt,bottom=1pt,colback  = white,colframe=myblue, boxrule=0.3mm]
\textbf{Summary:} Our analysis of Stack Overflow questions shows that users face challenges when configuring linter tools for coding standards.
\end{tcolorbox}

\section{Approach}\label{method}

\begin{figure*}
  \centering
    \includegraphics[width=5.5in]{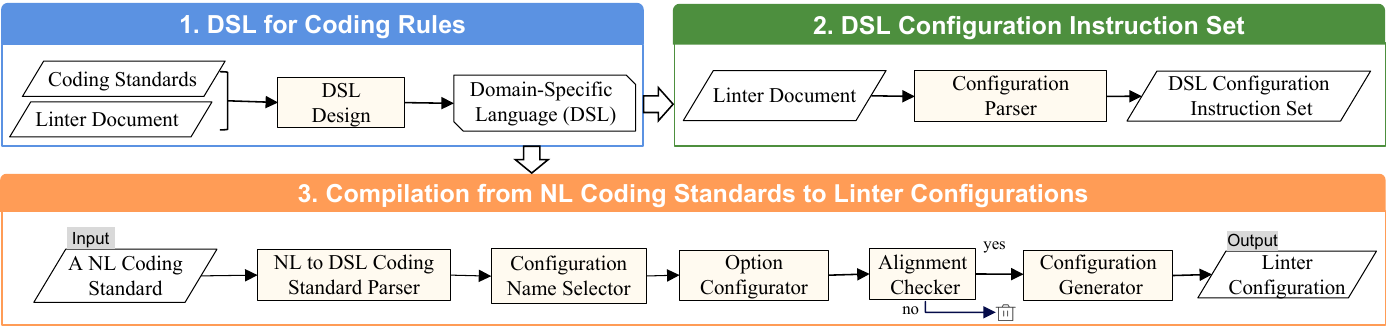}
     \caption{Approach overview of generating linter configurations for coding standards}
    \label{fig:app_overview} 
    \vspace{-3mm}
\end{figure*}

\begin{figure}
  \centering
    \setlength{\abovecaptionskip}{2mm}
    \includegraphics[width=5.5in]{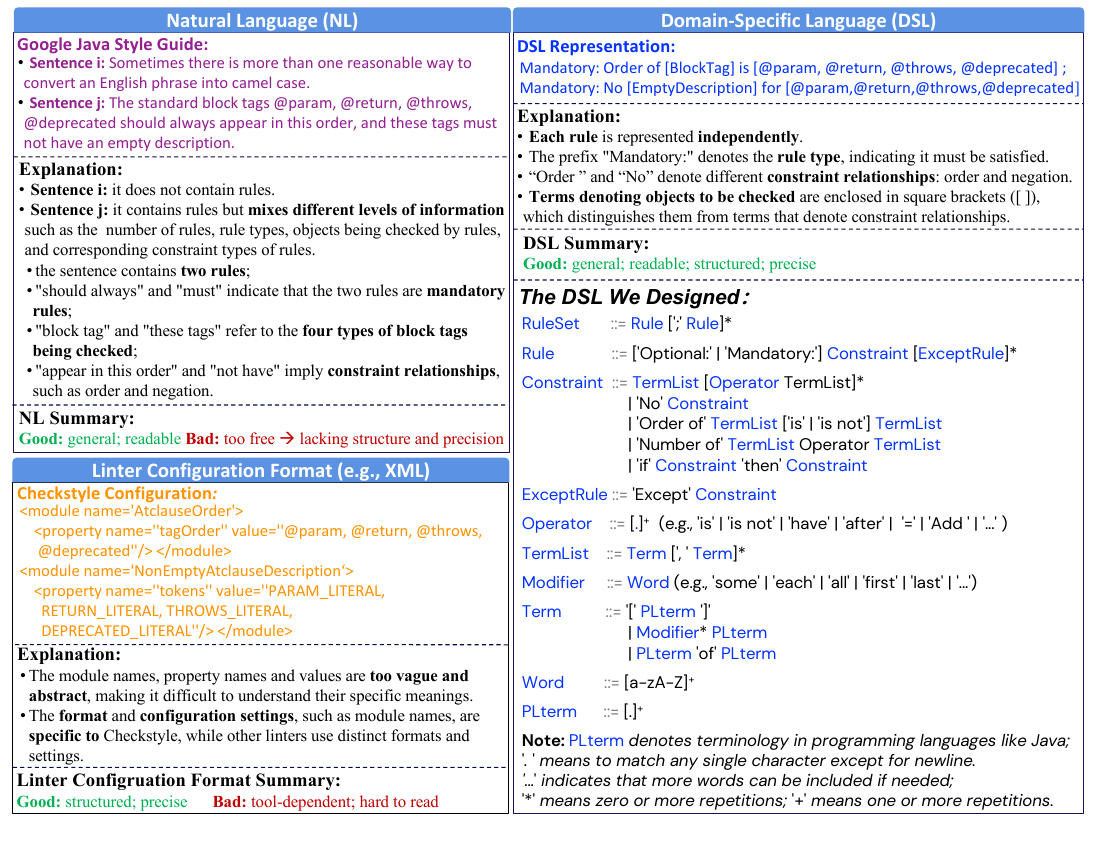}
    \caption{Comparison of coding rule representations (NL, linter configuration, and DSL)} 
    \label{fig:dsl_motivation} 
    \vspace{-3mm}
\end{figure}

Figure~\ref{fig:app_overview} illustrates our approach for generating linter configurations for coding standards, which consists of three steps. 
The first two steps are preparation phases that are generally performed once, while the third step is executed each time a coding standard needs to be configured in the linter. 
First, we design a generic domain-specific language (DSL) to clearly express the coding rules of both coding standards and linter configurations. 
Next, we build linter configurations as a DSL configuration instruction set by parsing linter documentation. 
Finally, for a given natural language (NL) coding standard, we employ a compilation pipeline to produce the corresponding linter configurations. 
This pipeline parses the NL coding standards into DSL coding standards, matches them with the DSL instruction set to determine configuration names, option names, and values, and verifies their alignment. 
Once aligned, the linter configurations are generated in tool-specific formats such as XML. 
Figure~\ref{fig:dsl_motivation} shows an example of NL coding standards and the corresponding Checkstyle configurations. 

\subsection{DSL for Coding Rules from Coding Standards and Linter Configurations}\label{sec: dsl_kb}

\subsubsection{Motivating Examples for DSL Design}~\label{sec:motivation_DSL} 
Similar to compiler design~\cite{louden1997compiler,grune2012modern,muchnick1997advanced}, which translates high-level code into machine-readable form through intermediate representations such as abstract syntax trees or bytecodes, our approach aims to compile NL coding standards into linter configurations. 
To support this, we design a DSL that expresses coding rules in a tool-agnostic, structured, precise, and readable form. 
While both NL coding standards and machine-readable linter configurations can represent coding rules, they do not meet the requirements. 

NL is tool-agnostic and readable but lacks structure and precision. 
A NL sentence of coding standards may or may not contain rules and can embed multiple layers of information, such as rule types, objects being checked, and constraint types. 
For instance, in Figure~\ref{fig:dsl_motivation}, the sentences in the \textit{NL box} illustrate this issue. 
In the Google Java coding standard, sentence $i$ does not have rules, while sentence $j$ has two rules. 
The two rules have mixed information: 
(1) ``should always'' and ``must'' indicate mandatory rules; 
(2) ``block tag'' and ``these tags'' refer to the four types of block tags being checked;
(3) ``appear in this order'' and ``not have'' imply order and negation constraint relationships. 

In contrast, machine-readable formats, such as XML, are structured and precise, yet tool-dependent and often too abstract to fully convey the intended semantics. 
In the \textit{Linter Configuration Format box} of Figure~\ref{fig:dsl_motivation}, the Checkstyle configuration for sentence $j$ consists of configuration names, option names and values. 
However, the configuration names, option names, and values are abstract and incomplete, not enough to convey the exact meaning. 
Moreover, linter configurations such as configuration names and formats are specific to Checkstyle and cannot be reused by other linters, which often use different settings and formats. 

To provide a tool-agnostic, structured, precise and readable representation for coding rules, we design the DSL shown in the \textit{domain-specific language} box from Figure~\ref{fig:dsl_motivation}.  
The corresponding DSL representations of sentence $j$ are ``\textcolor{gray}{\textit{Mandatory: Order of [BlockTag] is [@param, @return, @throws, @deprecated]}}'' and ``\textcolor{gray}{\textit{Mandatory: No [EmptyDescription] for [@param, @return, @throws, @deprecated]}}''. 
The DSL independently expresses each rule, clearly separating rule types (e.g., \textit{``Mandatory:''}), objects (enclosed in \textit{``[]''}), and constraint relationships (e.g., \textit{``Order of''} and \textit{``No''}).

\subsubsection{Information Source for Designing DSL}~\label{sec:source_DSL}
To design the DSL, we first select research objects of coding standards and linters. 
Java, with its mature ecosystem and rigorous coding practices~\cite{abdallah2017java}, represents a language with long-standing conventions, making it an ideal choice. 
Given the Google coding standards' broad acceptance, comprehensive coverage, and frequent reference across software projects and the industry~\cite{gleStyle}, we select the Google Java style guide as a representative coding standard. 
To enforce these standards, we select Checkstyle~\cite{ckstyle} as the linter for Java, as it is highly configurable, widely used, and has been extensively studied~\cite{yiu2023checkstyle,torunski2017code,loriot2022styler,edwards2017investigating}.
To understand linter configurations, we refer to Checkstyle documentation~\cite{doc_ckstyle}, which provides detailed information about linter configurations. 
We emphasize that our choice of the Google Java style guide and Checkstyle is not aimed at targeting these specific coding standards or linters. 
Rather, we select them due to their broad adoption, comprehensive coverage, and representativeness, making them ideal entry points for designing the DSL. 
\subsubsection{Process of Designing DSL} We use a card sorting approach~\cite{card_sorting} to define DSL by analyzing a sample of 320 sentences with a confidence level of 95\% and a confidence interval of 5 drawn from all sentences from both Google Java style guide and Checkstyle documentation.  
The process of designing DSL has two iterations and we used two evaluators. 
In the first iteration, we randomly sample 175 sentences with a confidence level of 95\% and a confidence interval of 5 from 320 sentences. 
 Two authors first independently represent each sentence using their designed DSLs, and then they discuss to construct a DSL. 
 In the second iteration, two of the authors independently represent remaining sentences with the DSL. 
  If the DSL is not enough to represent the rule of a sentence, they annotate the sentence with a brief description. 
  They find that there are no sentences that cannot be represented using the DSL. 
  We use Cohen's Kappa measure~\cite{kappa_value} to examine the agreement between the two authors. 
  The Kappa value is 0.7, which indicates a high agreement between two authors. 
  Finally, two authors discuss the disagreements to reach an agreement.

\subsubsection{DSL for Coding Standards and Linter Configurations}\label{dsl_grammar} 
Figure~\ref{fig:dsl_motivation} shows the designed DSL. 
We only provide necessary explanation of the DSL grammar below. 
A complete explanation of the DSL grammar can be found in \textcolor{blue}{\href{https://github.com/anynomousaccount/lint_gen_kod/blob/main/Appendices.pdf}{APPENDIX B}}.

\begin{itemize}[left=-1.3pt, labelsep=0.5mm,leftmargin=*,itemindent=1em,align=left]
\item \textit{RuleSet} represents sets of \textit{Rules}, where individual rules are separated by semicolons. 
\item A \textit{Rule} comprises a \textit{RuleType}, a \textit{Constraint}, and optionally any number of \textit{ExceptRule}. 
The \textit{RuleType} can be `Mandatory' or `Optional', indicating whether the rule must be followed or is optional. 
The \textit{ExceptRule} specifies exceptions to a \textit{Constraint}. 
 \textit{Constraint} has five types. 
 We explain the first two constraint relationships below, and the others are in \textcolor{blue}{\href{https://github.com/anynomousaccount/lint_gen_kod/blob/main/Appendices.pdf}{APPENDIX B}}.

\item \textit{TermList [Operator TermList]*}: Denotes the relationship that TermList must satisfy, expressed via operators. 
For example, ``Checks for braces around code blocks'' from the Checkstyle documentation is represented in the DSL as ``\textcolor{gray}{\textit{Mandatory: [CodeBlock] have [Brace]}}''.


\item \textit{`No' Constraint}: The \textit{Constraint} is prohibited. 
For example, in the Figure~\ref{fig:dsl_motivation}, the NL description of the Google Java coding standard, ``these tags must not have an empty
description'' is represented in the DSL as ``\textcolor{gray}{\textit{Mandatory: No [EmptyDescription] for [@param, @return, @throws, @deprecated]}}''.

\item \textit{Operator} denotes the relationship \textit{TermList}s should satisfy. 

\item \textit{TermList} is a list of \textit{Term}s separated by commas.




\item \textit{PLterm} represents terminology specific to programming languages, such as ``BlockTag'' and ``@param'' in the DSL representation of the Figure~\ref{fig:dsl_motivation}. 
It can be any sequence of characters, excluding newlines. 
\end{itemize}
\subsection{Building Linter Configurations into a DSL Configuration Instruction Set}\label{sec:tool_config_model}

\begin{figure}
  \centering
    \includegraphics[width=4.3in]{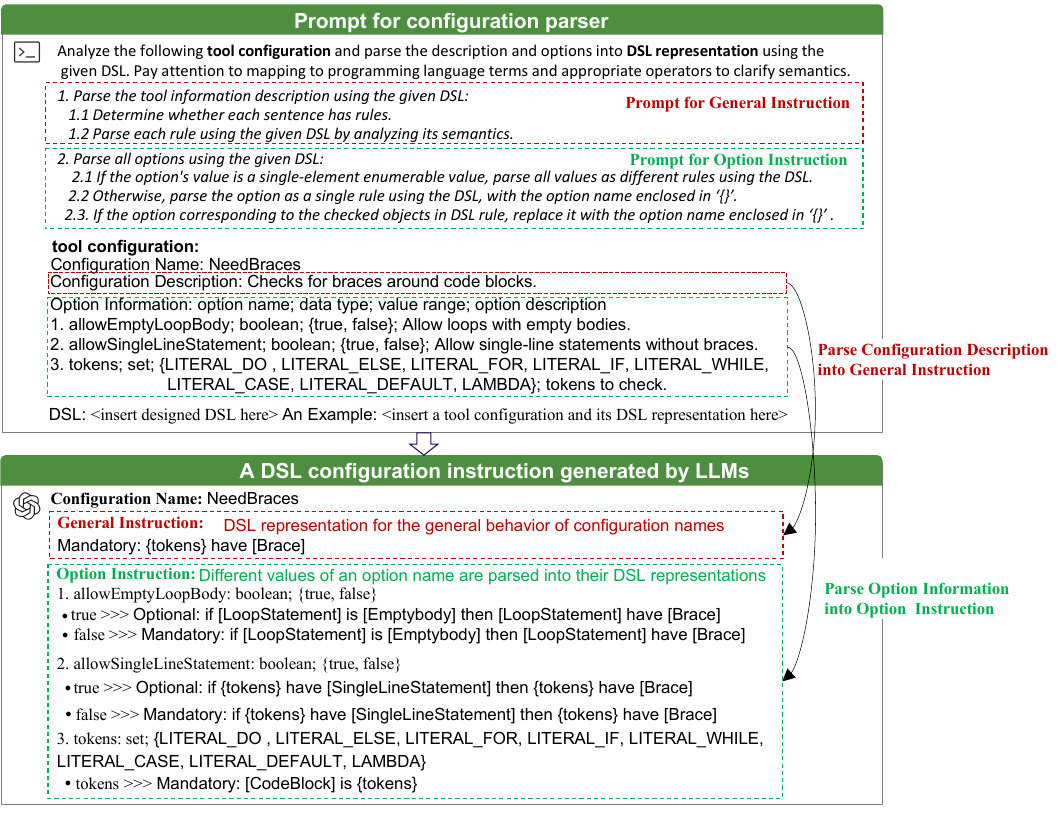}
    \caption{Building a linter configuration into the DSL configuration instruction from the linter document}
    \label{fig:tool_dsl} 
    \vspace{-0.4cm}
\end{figure}

A linter configuration consists of two key components: a configuration name and its associated options. 
The name specifies the general purpose of the configuration, and the options control its specific behaviors. 
Therefore, we parse each linter configuration into a DSL instruction comprising the general instruction and the option instruction. 


Building on the generative capabilities of LLMs, we prompt the LLM to parse a linter configuration into its general instructions and option-specific instructions in DSL. 
We provide an example to clarify the task and the expected response format. 
Figure~\ref{fig:tool_dsl} presents the prompt and LLM response for \textit{NeedBrace} configuration. 

The general instruction and option instruction are detailed below. 

\noindent \scalebox{1.5}{\textbullet} \textbf{General Instruction}: It denotes the general behavior of the configuration name. 
The configuration description summarizes the general functionality of the configuration name. 
Thus, we parse the configuration description into the general DSL instruction.  
Since the description often contains multiple sentences, some of which are code examples or explanations rather than rules, we first prompt the LLM to determine each sentence express coding rules, as shown in the step 1.1 of the prompt in Figure~\ref{fig:tool_dsl}. 
Once identified, we then prompt the LLM to parse each coding rule using the DSL, as shown in the step 1.2 of the prompt in Figure~\ref{fig:tool_dsl}.  

\noindent \scalebox{1.5}{\textbullet} \textbf{Option Instruction}: It denotes different behaviors for different option values of a option name. 
For each option, different option values correspond to different behaviors. 
Thus, we generate a separate DSL representation for each value to accurately capture its corresponding behavior. 
Specifically, for finite enumerated values (e.g., boolean values true and false), we parse different values into different DSL rules, as shown in the prompt of step 2.1 in Figure~\ref{fig:tool_dsl}. 
Otherwise, we parse the option into a single DSL rule by enclosing the option name in brackets, as shown in the prompt of step 2.2 in Figure~\ref{fig:tool_dsl}. 
This is because parsing each option value for infinite or numerous values is impractical and redundant, enclosed the option name in brackets allows for later option value assignment. 
We also observe that some options specify the objects to be checked. 
Similar to the infinite-value case, we replace these objects with the option name in brackets, as shown in the prompt of step 2.3 in Figure~\ref{fig:tool_dsl}. 
For example, the ``allowSingleLineStatement'' option only has two values: true and false. 
Therefore, we parse each value into a separate DSL rule. 
Since ``tokens'' option corresponds to the checked object ``CodeBlock'', we replace it with ``\{tokens\}'' for later option value assignment.







\subsection{Compilation from NL Coding Standards to Linter Configurations} 

Directly using a single prompt to generate linter configurations for coding standards is inefficient, as this approach can reduce scalability, complicate debugging, and can easily exceed LLM token limits due to the large number of linter configurations included in the prompt. 
Inspired by compiler design, which translates high-level representations into machine code through stages like parsing, intermediate code generation, semantic analysis, and machine code generation, we split the compilation into five steps, where each step only focus on a task to ensure the correctness and scalability. 
The DSL and the DSL configuration instruction set provide a foundation for compiling a NL coding standard into the linter configurations. 
For each step, we provide an example for LLMs to better understand the task. 

\begin{figure*}
  \centering
    \includegraphics[width=5.6in]{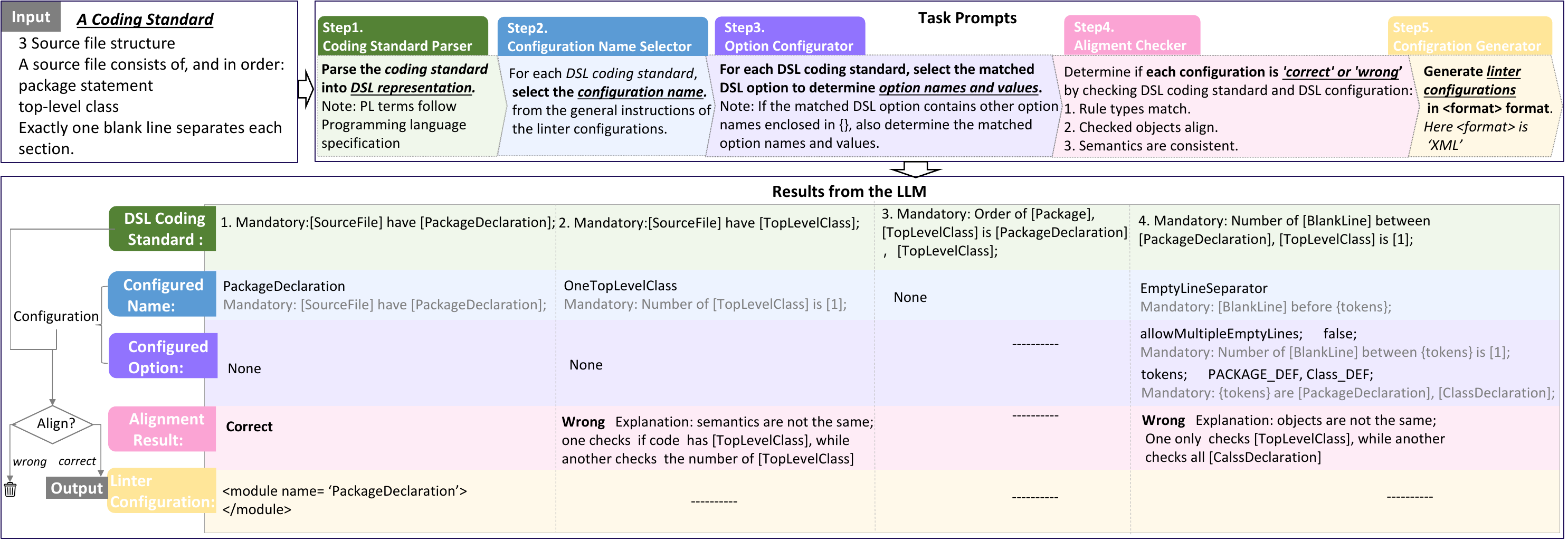}
    \caption{Compilation from a NL coding standard to linter configuration 
    }
    \label{fig:transpilation_tool_config_nl_cs} 
    \vspace{-0.5cm}
\end{figure*}

\subsubsection{NL to DSL Coding Standard Parser}~\label{sec:cs_dsl}
A natural language (NL) coding standard often contains non-rule sentences (e.g., examples or explanations) and may express multiple rules in a single sentence. 
To enable precise and structured interpretation, we parse the NL standard into the DSL rule representation as an intermediate representation for linter configuration generation. 
We prompt the LLM with a template comprising the task prompt, the NL coding standard, the DSL grammar from Section~\ref{dsl_grammar}, and an example pairing an NL coding standard with its DSL representation. 
Step 1 (green box) in Figure \ref{fig:transpilation_tool_config_nl_cs} illustrates the task prompt and the resulting DSL coding standard. 
For example, in Figure~\ref{fig:transpilation_tool_config_nl_cs}, the NL coding standard is parsed into the DSL coding standard comprising four DSL rules. 

\subsubsection{Configuration Name Selector} \label{sec:rule_selct} 
To generate configurations, it is natural to first identify the configuration name, followed by the specific options to be set. 
Directly inputting the entire DSL instruction set for name selection is unnecessary and may exceed the LLM's token limit. 
Since the general instruction captures the overall behavior of the configuration, matching it with the DSL coding standard is sufficient to identify candidate configuration names. 

We prompt the LLM with a template consisting of the task prompt, the DSL coding standard, general instructions of all linter configurations, and an example pairing the DSL coding standard with its corresponding configuration names. 
Step 2 in Figure~\ref{fig:transpilation_tool_config_nl_cs} shows the task prompt and the configuration names selected by the LLM. 
For example, for the first DSL coding standard in Figure~\ref{fig:transpilation_tool_config_nl_cs}, the LLM responds with ``PackageDeclaration'' configuration name from Checkstyle.  

\subsubsection{Option Configurator} \label{sec: gen_config} 
After selecting candidate configuration names for the DSL coding standard, we provide the LLM with the option instructions of configured names to determine the appropriate options. 
The LLM matches the DSL option with the DSL coding standard to identify relevant option names and values. 

The prompt template consists of the task prompt, the DSL coding standard, option instructions of the selected configuration names, and an example pairing the DSL coding standard with its configured options. 
The step 3 in Figure~\ref{fig:transpilation_tool_config_nl_cs} shows the task prompt and configured options from the LLM. 
For example, for the last DSL coding standard, 
the matching DSL option is ``\textcolor{gray}{\textit{Mandatory: Number of [BlankLine] between \{tokens\} is [1]}}''. 
Since the DSL option contains ``tokens'' option that enclosed in brackets, LLMs further extract the option name and set the value to ``PACKAGE\_DEF, CLASS\_DEF''.

\subsubsection{Alignment Checker}~\label{sec:config_validate} 
Correct configurations are essential, as incorrect ones can lead linters to report misleading issues, wasting developer time and potentially leading to wrong code changes. 
Meanwhile, we observe that LLMs tend to prioritize finding as many configurations as possible. 
To ensure correctness, we further verify the consistency between the generated DSL configuration and the DSL coding standard across three aspects: rule type consistency, object consistency, and semantic alignment. 

The prompt template consists of the task prompt, the DSL coding standard, the corresponding DSL configurations from general and option instructions, and an example pairing the DSL coding standard with its DSL configurations and alignment results. 
Step 4 of Figure~\ref{fig:transpilation_tool_config_nl_cs} shows the task prompt and the alignment results generated by the LLM. 
For example, for the last DSL coding standard, the checked objects is not the same as the checked objects of the corresponding DSL option, so the configuration is wrong. 
\subsubsection{Linter-Specific Configuration Generator} 
After obtaining aligned configurations, we convert them into linter-specific formats. 
We use LLMs for their flexibility and ability to adapt across linters with varying schemas, avoiding the need to maintain separate generators. 

The prompt template consists of task prompt, successfully verified configurations in Section~\ref{sec:config_validate}, an example pairing with verified configurations, and linter-specific configurations.   
Step 5 of Figure~\ref{fig:transpilation_tool_config_nl_cs} shows the task prompt and the linter-specific configurations. 
For example, only the first configuration is correct, so the final configuration uses the module ``PackageDeclaration''.
\section{Evaluation}\label{ourevaluation}
To evaluate our approach, we study five research questions:

\begin{enumerate}[fullwidth,itemindent=0em,leftmargin = 0pt]
\item[\textbf{RQ1:}] How effective are the DSL and the LLM in generating DSL representations for the Java coding standards and Checkstyle configurations? 

\item[\textbf{RQ2:}] 
How effective is our approach in generating Checkstyle configurations for the Java coding standards? 
\item[\textbf{RQ3:}] How do the different modules in our approach impact its effectiveness?

\item[\textbf{RQ4:}] How useful is our approach for developers in generating Checkstyle configurations for Java coding standards?

\item[\textbf{RQ5:}] Can our approach be effectively extended to other programming languages, coding standards, and linters? 
\end{enumerate}

\subsection{RQ1: Effectiveness of DSL Representation Generation}\label{rq_dsl}

\subsubsection{Motivation}\label{rq_dsl_motiv} 
After we design DSL in Section~\ref{sec: dsl_kb}, it is important to verify if the DSL and LLMs can effectively express coding standards and linter configurations. 

\subsubsection{Approach}\label{rq_dsl_approach} 
We manually review the correctness of DSL representations for Google Java coding standards and Checkstyle configurations. 
If the total number of coding standards or linter configurations is fewer than 100, we examine all instances; otherwise, we apply random sampling with a 95\% confidence level and a 5\% margin of error~\cite{singh2013elements}. 
For coding standards or linter configurations, two authors and two external experts, each with over six years of Java experience, independently classify each DSL representation as correct (a true positive), incorrect (a false positive), or missing (a false negative). 
We denote the number of true positives, false positives and false negatives as $TP$,  $FP$ and $FN$. 
Since a coding standard or linter configuration typically involves multiple DSL representations, they further evaluate whether all DSL representations for a given coding standard or linter configuration are correct. 
Any inconsistencies are resolved through discussion. 
We employ four metrics: precision, recall, F1-score and accuracy. 
Precision is calculated as $P = \frac{TP}{TP+FP}$, recall as $R = \frac{TP}{TP+FN}$, and F1-score as $F1 = \frac{2 \times P \times R}{P+R}$. 
Accuracy evaluates the proportion of coding standards or linter configurations for which all DSL rules are correctly generated. 
It is calculated as $Acc = \frac{\text{number of coding standards or linter configurations with correct DSL representation}}{\text{total number of coding standards or linter configurations}}$. 
We employ the widely adopted state-of-the-art GPT-4o~\cite{chat_gpt}, whose  performance and wide use in recent research make it an ideal prototype engine for our approach~\cite{wan2025divide,wang2025automating,yigao_test_refactor,yu2025cxxcrafter,xia2025demystifying}. 
To minimize output variability of the LLM, the temperature is 0.  

\begin{table}
\caption{Results of DSL generation for Google Java coding standards and Checkstyle configurations. 
}
\vspace{-0.3cm}
\label{tab:metric_dsl}
\centering
\scalebox{0.8}{
\begin{tabular}{llllllll}
\multicolumn{1}{c|}{\cellcolor{aliceblue}Name}&
\multicolumn{1}{c}{\cellcolor{aliceblue}\#Num}&
\multicolumn{1}{c}{\cellcolor{aliceblue}\#Samp}&
\multicolumn{1}{c}{\cellcolor{aliceblue}\#DSL}&
\multicolumn{1}{c}{\cellcolor{aliceblue}Acc(\%)}&
\multicolumn{1}{c}{\cellcolor{aliceblue}P(\%)}&
\multicolumn{1}{c}{\cellcolor{aliceblue}R(\%)}&
\multicolumn{1}{c}{\cellcolor{aliceblue}F1(\%)}\\ \hline

\multicolumn{1}{c|}{Java Style Guide}&
\multicolumn{1}{c}{68}&
\multicolumn{1}{c}{68}&
\multicolumn{1}{c}{182}&
\multicolumn{1}{c}{89.7}&
\multicolumn{1}{c}{96.2}&
\multicolumn{1}{c}{100}&
\multicolumn{1}{c}{98.0}\\
\hline

\multicolumn{1}{c|}{Checkstyle}&
\multicolumn{1}{c}{184}&
\multicolumn{1}{c}{125}&
\multicolumn{1}{c}{620}&
\multicolumn{1}{c}{83.2}&
\multicolumn{1}{c}{92.3}&
\multicolumn{1}{c}{100}&
\multicolumn{1}{c}{96.0}\\
\hline
\multicolumn{8}{p{300pt}}{\footnotesize Note: \#Num is the total number of coding standards or linter configurations. \#Samp is the number of sampled coding standards or linter configurations. \#DSL is the number of corresponding DSL representations generated by our approach. }\\%

\end{tabular}
}
\vspace{-5mm}
\end{table}
\subsubsection{Result}\label{rq_dsl_approach_res} 
Table~\ref{tab:metric_dsl} presents the accuracy, precision, recall, and F1-score for DSL representations of the Google Java coding standard and Checkstyle configurations. 
In total, we review 182 DSL representations across 68 coding standards and 620 DSL representations for 125 linter configurations. 
For the Google Java style guide and Checkstyle, the precision, recall, and F1 scores exceed 90\%, with accuracy above 80\%. 
The results demonstrate that the designed DSL and the LLM can effectively represent a wide range of coding rules in both Java coding standards and Checkstyle configurations. 

\noindent \textbf{Failure analysis of our approach}. By manually analyzing wrong and missing DSL representations for coding standards and linter configurations, we summarize two reasons as follows: 

\noindent (1) Incomplete information from a sentence in coding standards or linter documentation can lead to incorrect DSL representations. 
For instance, in the Checkstyle configuration for \href{https://checkstyle.sourceforge.io/version/10.13.0/checks/coding/fallthrough.html#FallThrough}{``FallThrough''}, the description states, ``The check honors special comments to suppress the warning, by default, the texts fallthru, fall thru, ...''; however, this description lacks complete information about which type of statements the comments apply to. 
In reality, the rule of Checkstyle pertains only to case blocks, but the DSL instruction set builder incorrectly parses the sentence as 
\textcolor{gray}{\textit{``Optional: [Comment] matches [fallthru, fall thru, ...] suppress [Warning]''}}, omitting the specification that the ``[Comment] of [CaseBlock]''. 

\noindent (2) DSL parsers for coding standards or linter configurations may misinterpret whether a sentence defines rules, leading to missed or incorrect DSL representations. 
For example, \href{https://google.github.io/styleguide/javaguide.html#s4.4-column-limit}{the NL description} in the Google Java coding standard, ``A character means any Unicode code point'' is an explanatory sentence instead of a rule. 
However, the DSL parser mistakenly interprets and parses it as a rule.

\vspace{-2mm}
\begin{tcolorbox}[width=5.5in,boxsep=1pt,left=1pt,right=1pt,top=1pt,bottom=1pt, colback=white,colframe=myblue, boxrule=0.3mm]
\textbf{Summary:} The high accuracy, F1-score, precison and recall shows the DSL and LLMs can effectively express Java coding standards and Checkstyle configurations into DSL representations. 
\end{tcolorbox}

\subsection{RQ2: Effectiveness of Linter Configuration Generation}

\subsubsection{Motivation}\label{rq1_approa} 
The success of ChatGPT~\cite{intro_chat_gpt} demonstrates the ability of LLMs to comprehend prompts and complete tasks. 
We aim to compare our approach with existing LLM-based methods, highlighting how our approach effectively generating linter configurations for coding standards.



\begin{table}
\caption{Benchmark statistics for Checkstyle configurations on Google Java coding standard.} 
\vspace{-2mm}
\label{tab:benchmark}
\centering
\scalebox{0.8}{
\begin{tabular}{ll}
\multicolumn{1}{c|}{\cellcolor{aliceblue} Configuration Category for Coding Standard}&
\multicolumn{1}{c}{\cellcolor{aliceblue}Number}\\ 
\hline
\multicolumn{1}{l|}{No Configuration} &
\multicolumn{1}{c}{19} \\\hline

\multicolumn{1}{l|}{Has Configuration} &
\multicolumn{1}{c}{49} \\\hline

\multicolumn{1}{l|}{Config with Only Configuration Name} &
\multicolumn{1}{c}{7}\\ \hline

\multicolumn{1}{l|}{Config with Configuration Name and Options} &
\multicolumn{1}{c}{42} \\ \hline

\multicolumn{1}{l|}{Config with Multiple Configurations} &
\multicolumn{1}{c}{13} \\ \hline

\multicolumn{1}{l|}{Total Coding Standards} &
\multicolumn{1}{c}{68} \\ 
\hline

\end{tabular}
}
\vspace{-3mm}
\end{table}

\begin{figure}
  \centering
    \includegraphics[width=4.6in]{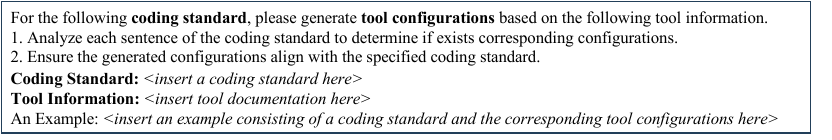}
    \vspace{-0.4cm}
    \caption{Prompt template of baselines}
    \label{prompt_template_baselines} 
    \vspace{-0.6cm}
\end{figure}

\subsubsection{Approach}\label{rq1_approa} 

\noindent \textbf{Benchmark.} 
To evaluate the effectiveness of our approach, we create a benchmark of linter configurations for coding standards in the form of $<$coding standard, tool configuration$>$. 
Since no existing dataset is available, we manually construct Checkstyle configurations for the Google Java coding standard using 120 hours. 
Two authors and two external experts, each with over six years of Java programming experience, independently create these configurations and resolve any inconsistencies through discussion. 
Table~\ref{tab:benchmark} presents the statistics for the benchmark. 
For the Google Java coding standard, 28\% of the 68 coding standards lacked configurations, while 72\% had configurations. 
Among those with configurations, 10\% contained only configuration names, 62\% included both names and options, and 19\% involved multiple configuration names. 
The result highlights the challenge in generating linter configurations, as many coding standards either lack corresponding configurations or require complex configurations with options or with multiple configuration names. 

\noindent \textbf{Baselines.} 
Since linter documentation contains information at different levels, and Retrieval-Augmented Generation (RAG)~\cite{gao2023retrieval} is an effective method for enhancing LLM capabilities, we design six baselines. 
The prompt is as shown in Figure~\ref{prompt_template_baselines}. 
To determine the prompt, two authors and two external experts independently review the coding standards and linter configurations to draft the prompt. 
They then discuss their results to finalize the best  prompt. 
Here we provide an example which is aligned with the example in our approach to maintain fairness.  
The LLM settings are consistent with those described in Section~\ref{rq_dsl_approach}. 

\vspace{-1mm}
\begin{enumerate}[label=\arabic*., left=-1pt, labelsep=1mm,leftmargin=*,labelindent=0pt]
\item \textbf{Closed Book} (i.e., $<$tool information$>$ is empty): The baseline directly uses the LLM to generate linter configurations for coding standards without providing additional context or knowledge. 
The $<$tool information$>$ of the prompt template is empty.  
It serves as a fundamental comparison point to assess the impact of incorporating additional information on effectiveness of our approach.

\item \textbf{Name} (i.e., $<$tool information$>$ consists only of linter configuration names): To augment LLM's capabilities, leveraging additional information or knowledge with LLMs is a well-established practice~\cite{ren2023api,martino2023knowledge,ding2023integrating,zhang2024refactoring}. 
Linters for coding standards generally provide official documentation~\cite{doc_ckstyle} that includes configuration names, descriptions, and option information. 
Given that varying levels of granularity in the information might affect the performance of LLMs, we explore different baselines by utilizing LLMs with varying levels of tool information granularity. 
 In this baseline, the LLM generates configurations for coding standards using only the tool configuration names.

\item \textbf{Name+Desc} (i.e., $<$tool information$>$ consists of linter configuration names and configuration descriptions): It uses the LLM with both the linter configuration names and their descriptions to generate configurations for coding standards. 
\item \textbf{Name+Desc+Opts} (i.e., $<$tool information$>$ consists of linter configuration names, descriptions and option information): It uses the LLM with the linter configuration names, their descriptions, and information of options to generate configurations for coding standards. 
\item \textbf{RAG (Name+Desc)} (i.e., Retrieval-Augmented Generation with tool configuration names and descriptions): 
RAG is a method that enhances generation models by first retrieving relevant knowledge from knowledge bases and then integrating it into the generation process~\cite{gao2023retrieval}.  
RAG has been widely used in several software engineering tasks~\cite{gao2023retrieval,zhao2024retrieval,ding2024survey,mackay2024test}.  

For this baseline, we use official linter documentation, including configuration names and descriptions, as external knowledge. 
In the indexing phase, we encode this external knowledge into vector representations using the OpenAI embedding model ada-002~\cite{openai_embed}, and store these vectors in ChromaDB. 
During retrieval, we query the top-k semantically similar linter configuration names for a given coding standard. 
According to Jiang et al.~\cite{jiang2024longrag}, we set k to 10, considering that a higher k value can overwhelm models and hinder information extraction in long contexts. 
In the generation phase, we integrate the retrieved top-k linter configuration into the LLM to generate configurations for coding standards. 
The $<$tool information$>$ of the prompt template refers to the configuration names and corresponding descriptions of the retrieved top-k linter configurations. 

\item \textbf{RAG (Name+Desc+Opts)} (i.e., Retrieval-Augmented Generation with linter configuration names, descriptions, and option information): It follows the same process as the fifth baseline. 
The one difference is the inclusion of RAG, which leverages linter configuration names, descriptions, and option details as external knowledge, in conjunction with LLMs, to generate configurations for coding standards.
\end{enumerate}

\noindent \textbf{Metrics.} 
For a coding standard, linter configuration typically includes the configuration name, option name and the corresponding option values. 
Consequently, for one metric, we evaluate effectiveness at three levels of granularity: configuration name, option name, and option value, as shown in Table~\ref{tab:metrics_res}.

We first use the accuracy metric, which measures the percentage of coding standards for which the approach correctly generates linter configurations. 
The formula of the accuracy of an approach  is $Acc=\frac{\text{number of coding standards with correct configurations}}{\text{number of coding standards}}$. 
Since a coding standard may have multiple linter configurations, we further refine our evaluation metrics. 
For a coding standard, a true positive occurs when a configuration given by an approach exists in the benchmark. 
A false positive occurs when a configuration given by an approach does not exist in the benchmark. 
A false negative occurs when a configuration not given by an approach exists in the benchmark. 
We denote the number of true positives, false positives and false
negatives as $TP$, $FP$ and $FN$. 
We calculate the precision as $P = \frac{TP}{TP+FP}$, the recall as $R = \frac{TP}{TP+FN}$, and the F1-score  as $F1 = \frac{2\times P \times R}{P+R}$. 
The $Acc$ is measured at the level of all configurations for a coding standard, and $P$, $R$, and $F1$ are measured at the level of each configuration for a coding standard.

\begin{table*}
\centering
\caption{Results of generating Checkstyle configurations for Google Java coding standards. 
} 
\vspace{-2mm}
\label{tab:metrics_res}
\centering
\scalebox{0.74}{
\begin{tabular}{c|llllllllllll|}

\hline
\cellcolor{aliceblue}
&
\multicolumn{4}{c}{\cellcolor{aliceblue}Config Name Level Metrics (\%)} &
\multicolumn{4}{|c}{\cellcolor{aliceblue}Option Name Level Metrics(\%)} &
\multicolumn{4}{|c}{\cellcolor{aliceblue}Option Value Level Metrics(\%)} \\
\cline{2-13}\multirow{-2}{*}{{\cellcolor{aliceblue}}Approach} &
\multicolumn{1}{c}{\cellcolor{aliceblue}Acc} &
\multicolumn{1}{c}{\cellcolor{aliceblue}P} &
\multicolumn{1}{c}{\cellcolor{aliceblue}R} &
\multicolumn{1}{c}{\cellcolor{aliceblue}F1} &
\multicolumn{1}{|c}{\cellcolor{aliceblue}Acc} &
\multicolumn{1}{c}{\cellcolor{aliceblue}P} &
\multicolumn{1}{c}{\cellcolor{aliceblue}R} &
\multicolumn{1}{c}{\cellcolor{aliceblue}F1} &
\multicolumn{1}{|c}{\cellcolor{aliceblue}Acc} &
\multicolumn{1}{c}{\cellcolor{aliceblue}P} &
\multicolumn{1}{c}{\cellcolor{aliceblue}R} &
\multicolumn{1}{c}{\cellcolor{aliceblue}F1} 
\\ \hline

\multicolumn{1}{l|}{Closed-book}&\multicolumn{1}{c}{32.4}&\multicolumn{1}{c}{\underline{39.4}}&\multicolumn{1}{c}{55.4}&\multicolumn{1}{c}{\underline{46.1}}&\multicolumn{1}{|c}{19.1}&\multicolumn{1}{c}{26.0}&\multicolumn{1}{c}{36.5}&\multicolumn{1}{c}{30.3}&\multicolumn{1}{|c}{32.4}&\multicolumn{1}{c}{\underline{29.8}}&\multicolumn{1}{c}{41.9}&\multicolumn{1}{c}{\underline{34.8}}\\ \hline

\multicolumn{1}{l|}{Name}&\multicolumn{1}{c}{39.7}&\multicolumn{1}{c}{16.8}&\multicolumn{1}{c}{74.3}&\multicolumn{1}{c}{27.4}&\multicolumn{1}{|c}{23.5}&\multicolumn{1}{c}{11.6}&\multicolumn{1}{c}{51.4}&\multicolumn{1}{c}{19.0}&\multicolumn{1}{|c}{32.4}&\multicolumn{1}{c}{12.8}&\multicolumn{1}{c}{56.8}&\multicolumn{1}{c}{20.9}\\ \hline

\multicolumn{1}{l|}{Name$+$Desc}&\multicolumn{1}{c}{39.7}&\multicolumn{1}{c}{26.2}&\multicolumn{1}{c}{75.7}&\multicolumn{1}{c}{38.9}&\multicolumn{1}{|c}{29.4}&\multicolumn{1}{c}{21.0}&\multicolumn{1}{c}{60.8}&\multicolumn{1}{c}{31.2}&\multicolumn{1}{|c}{29.4}&\multicolumn{1}{c}{18.7}&\multicolumn{1}{c}{54.1}&\multicolumn{1}{c}{27.8}\\ \hline

\multicolumn{1}{l|}{Name$+$Desc$+$Opts}&\multicolumn{1}{c}{\underline{48.5}}&\multicolumn{1}{c}{23.4}&\multicolumn{1}{c}{\underline{78.4}}&\multicolumn{1}{c}{36.0}&\multicolumn{1}{|c}{\underline{48.5}}&\multicolumn{1}{c}{23.0}&\multicolumn{1}{c}{\underline{77.0}}&\multicolumn{1}{c}{35.4}&\multicolumn{1}{|c}{\underline{38.2}}&\multicolumn{1}{c}{19.0}&\multicolumn{1}{c}{\underline{63.5}}&\multicolumn{1}{c}{29.2}\\ \hline

\multicolumn{1}{l|}{RAG (Name$+$Desc)}&\multicolumn{1}{c}{14.7}&\multicolumn{1}{c}{24.4}&\multicolumn{1}{c}{73.0}&\multicolumn{1}{c}{36.6}&\multicolumn{1}{|c}{13.2}&\multicolumn{1}{c}{21.7}&\multicolumn{1}{c}{64.9}&\multicolumn{1}{c}{32.5}&\multicolumn{1}{|c}{13.2}&\multicolumn{1}{c}{19.0}&\multicolumn{1}{c}{56.8}&\multicolumn{1}{c}{28.5}\\ \hline

\multicolumn{1}{l|}{RAG (Name$+$Desc$+$Opts)}&\multicolumn{1}{c}{25.0}&\multicolumn{1}{c}{31.8}&\multicolumn{1}{c}{73.0}&\multicolumn{1}{c}{44.3}&\multicolumn{1}{|c}{25.0}&\multicolumn{1}{c}{\underline{31.8}}&\multicolumn{1}{c}{73.0}&\multicolumn{1}{c}{\underline{44.3}}&\multicolumn{1}{|c}{13.2}&\multicolumn{1}{c}{18.8}&\multicolumn{1}{c}{43.2}&\multicolumn{1}{c}{26.2}\\ \hline

\multicolumn{1}{l|}{\textbf{Our Method}}&\multicolumn{1}{c}{\textbf{73.5}}&\multicolumn{1}{c}{\textbf{81.3}}&\multicolumn{1}{c}{\textbf{82.4}}&\multicolumn{1}{c}{\textbf{81.9}}&\multicolumn{1}{|c}{\textbf{73.5}}&\multicolumn{1}{c}{\textbf{81.3}}&\multicolumn{1}{c}{\textbf{82.4}}&\multicolumn{1}{c}{\textbf{81.9}}&\multicolumn{1}{|c}{\textbf{72.1}}&\multicolumn{1}{c}{\textbf{69.3}}&\multicolumn{1}{c}{\textbf{70.3}}&\multicolumn{1}{c}{\textbf{69.8}}\\ \hline

\multicolumn{1}{l|}{Change (\%)}&\multicolumn{1}{c}{$\uparrow$51.5}&\multicolumn{1}{c}{$\uparrow$106.3}&\multicolumn{1}{c}{$\uparrow$5.1}&\multicolumn{1}{c}{$\uparrow$77.7}&\multicolumn{1}{|c}{$\uparrow$51.5}&\multicolumn{1}{c}{$\uparrow$155.7}&\multicolumn{1}{c}{$\uparrow$7.0}&\multicolumn{1}{c}{$\uparrow$84.9}&\multicolumn{1}{|c}{$\uparrow$88.7}&\multicolumn{1}{c}{$\uparrow$132.6}&\multicolumn{1}{c}{$\uparrow$10.7}&\multicolumn{1}{c}{$\uparrow$100.6}\\ \hline

\multicolumn{13}{p{500pt}}{Note: Bold values represent the best performance across all approaches. Underlined values represent the best performance among the baselines. ``Change'' shows the minimum improvement over the baselines. }\\ 

\end{tabular}
}
\vspace{-5mm}
\end{table*}
\subsubsection{Result}~\label{sec:res_gen_config_metric} 
Table~\ref{tab:metrics_res} presents the accuracy, precision, recall, and F1-score of our approach and the baselines. 
Our approach achieves 72.1\%$\sim$73.5\% for accuracy, 69.3\%$\sim$81.3\% for precision, 70.3\%$\sim$82.4\% for recall, and 69.8\%$\sim$81.9\% for F1-score across three levels of granularity. 
Our approach consistently outperforms the baselines in accuracy, precision, recall, and F1-score across the three levels of granularity, with minimum improvements of 51.5\%, 106.3\%, 5.1\%, and 77.7\%, respectively. 
Among the baselines, precision is notably low, indicating their tendency to generate incorrect configurations.
Even with traditional augmentation methods, such as incorporating tool documentation or the RAG strategy, precision remains low and may even degrade. 
In contrast, our approach significantly improves precision (by over 100\%) while maintaining strong recall, showing its effectiveness in generating accurate linter configurations. 

\noindent \textbf{Failure analysis of our approach.} To explore why our approach cannot generate configurations for coding standards, we examine configurations in the benchmark but our approach does not generate. 
The primary reason is that accurate linter configuration generation may require strong semantic inference and domain knowledge, which our approach lacks. 
For example, the Java coding standard for a summary fragment, ``...It does not begin with `A{@code Foo}...' '', is represented in DSL as \textcolor{gray}{\textit{``Mandatory: No [SummaryFragment] begin with [A{@code Foo}...]''}}. 
It should map to the ``forbiddenSummaryFragments'' option in the \href{https://checkstyle.sourceforge.io/version/10.13.0/checks/javadoc/summaryjavadoc.html#SummaryJavadoc}{SummaryJavadoc} configuration, represented in DSL as \textcolor{gray}{\textit{``Mandatory: [JavadocSummary] not have \{forbiddenSummaryFragments\}''}}. 
The option is a regular expression that can be set to match strings starting with ``[A{@code Foo}...]''. 
However, our DSL instruction configurator fails to recognize it due to limited reasoning capability.
\begin{tcolorbox}[width=5.5in,boxsep=1pt,left=1pt,right=1pt,top=1pt,bottom=1pt,colback  = white,colframe=myblue,, boxrule=0.3mm]
\textbf{Summary:} Our approach achieves superior accuracy, F1-score, precision, and recall compared to other baselines in generating Checkstyle configurations for Java coding standards, particularly at the fine-grained level and in terms of precision.
\end{tcolorbox}
\vspace{-1mm}

\subsection{RQ3: Ablation Study}\label{ablationstudy}
\subsubsection{Motivation} 
The effectiveness of our approach motivates us to assess the contribution of each module involved in the DSL-driven and AI chain compilation approach. 
\subsubsection{Approach} We conduct an ablation study with three variants based on our approach. 
The LLM settings are the same as those described in Section~\ref{rq1_approa}.

\noindent \textbf{w/o DSL:} Disables the generation of DSL representations (parsers) for coding standards and linter configurations. 

\noindent \textbf{w/o Selector:} Skips the configuration name selection step and directly generates configuration names and options.

\noindent \textbf{w/o Checker:} Omits the alignment checker that verifies whether the generated configurations conform to the coding standards.

\begin{table*}
\centering
\caption{Results of ablation study. 
} 
\vspace{-2mm}
\label{tab:ablation}
\centering
 \scalebox{0.7}{
\begin{tabular}{c|llllllllllll|}

\hline
\cellcolor{aliceblue}
&
\multicolumn{4}{c}{\cellcolor{aliceblue}Config Name Level Metrics (\%)} &
\multicolumn{4}{|c}{\cellcolor{aliceblue}Option Name Level Metrics(\%)} &
\multicolumn{4}{|c}{\cellcolor{aliceblue}Option Value Level Metrics(\%)} \\
\cline{2-13} \multirow{-2}{*}{{\cellcolor{aliceblue}}Approach} &
\multicolumn{1}{c}{\cellcolor{aliceblue}Acc} &
\multicolumn{1}{c}{\cellcolor{aliceblue}P} &
\multicolumn{1}{c}{\cellcolor{aliceblue}R} &
\multicolumn{1}{c}{\cellcolor{aliceblue}F1} &
\multicolumn{1}{|c}{\cellcolor{aliceblue}Acc} &
\multicolumn{1}{c}{\cellcolor{aliceblue}P} &
\multicolumn{1}{c}{\cellcolor{aliceblue}R} &
\multicolumn{1}{c}{\cellcolor{aliceblue}F1} &
\multicolumn{1}{|c}{\cellcolor{aliceblue}Acc} &
\multicolumn{1}{c}{\cellcolor{aliceblue}P} &
\multicolumn{1}{c}{\cellcolor{aliceblue}R} &
\multicolumn{1}{c}{\cellcolor{aliceblue}F1} 
\\ \hline


\multicolumn{1}{l|}{\textbf{Our Method}}&\multicolumn{1}{c}{\textbf{73.5}}&\multicolumn{1}{c}{\textbf{81.3}}&\multicolumn{1}{c}{82.4}&\multicolumn{1}{c}{\textbf{81.9}}&\multicolumn{1}{|c}{\textbf{73.5}}&\multicolumn{1}{c}{\textbf{81.3}}&\multicolumn{1}{c}{82.4}&\multicolumn{1}{c}{\textbf{81.9}}&\multicolumn{1}{|c}{\textbf{72.1}}&\multicolumn{1}{c}{\textbf{69.3}}&\multicolumn{1}{c}{70.3}&\multicolumn{1}{c}{\textbf{69.8}}\\ \hline
\hline

\multicolumn{1}{l|}{w/o DSL}&\multicolumn{1}{c}{61.8}&\multicolumn{1}{c}{67.0}&\multicolumn{1}{c}{76.7}&\multicolumn{1}{c}{71.5}&\multicolumn{1}{|c}{60.3}&\multicolumn{1}{c}{65.3}&\multicolumn{1}{c}{75.6}&\multicolumn{1}{c}{70.1}&\multicolumn{1}{|c}{51.5}&\multicolumn{1}{c}{52.2}&\multicolumn{1}{c}{60.2}&\multicolumn{1}{c}{55.9}\\ 
\multicolumn{1}{l|}{(\textit{Change})}&\multicolumn{1}{c}{($\downarrow$18.9)}&\multicolumn{1}{c}{($\downarrow$21.3)}&\multicolumn{1}{c}{($\downarrow$7.4)}&\multicolumn{1}{c}{($\downarrow$14.5)}&\multicolumn{1}{|c}{($\downarrow$21.9)}&\multicolumn{1}{c}{($\downarrow$24.5)}&\multicolumn{1}{c}{($\downarrow$9.0)}&\multicolumn{1}{c}{($\downarrow$16.8)}&\multicolumn{1}{|c}{($\downarrow$40.0)}&\multicolumn{1}{c}{($\downarrow$32.8)}&\multicolumn{1}{c}{($\downarrow$16.8)}&\multicolumn{1}{c}{($\downarrow$24.9)}\\ 
\hline
\hline

\multicolumn{1}{l|}{w/o Selector}&\multicolumn{1}{c}{60.3}&\multicolumn{1}{c}{72.9}&\multicolumn{1}{c}{47.3}&\multicolumn{1}{c}{57.4}&\multicolumn{1}{|c}{58.9}&\multicolumn{1}{c}{70.8}&\multicolumn{1}{c}{45.9}&\multicolumn{1}{c}{55.7}&\multicolumn{1}{|c}{57.3}&\multicolumn{1}{c}{66.7}&\multicolumn{1}{c}{43.2}&\multicolumn{1}{c}{52.5}\\ 
\multicolumn{1}{l|}{(\textit{Change})}&\multicolumn{1}{c}{($\downarrow$21.9)}&\multicolumn{1}{c}{($\downarrow$11.5)}&\multicolumn{1}{c}{($\downarrow$74.2)}&\multicolumn{1}{c}{($\downarrow$42.7)}&\multicolumn{1}{|c}{($\downarrow$24.8)}&\multicolumn{1}{c}{($\downarrow$14.8)}&\multicolumn{1}{c}{($\downarrow$79.5)}&\multicolumn{1}{c}{($\downarrow$47.0)}&\multicolumn{1}{|c}{($\downarrow$25.8)}&\multicolumn{1}{c}{($\downarrow$3.9)}&\multicolumn{1}{c}{($\downarrow$62.7)}&\multicolumn{1}{c}{($\downarrow$33.0)}\\ 
\hline
\hline

\multicolumn{1}{l|}{w/o Checker}&\multicolumn{1}{c}{73.5}&\multicolumn{1}{c}{75.0}&\multicolumn{1}{c}{\textbf{82.9}}&\multicolumn{1}{c}{78.8}&\multicolumn{1}{|c}{72.1}&\multicolumn{1}{c}{73.8}&\multicolumn{1}{c}{\textbf{82.9}}&\multicolumn{1}{c}{78.1}&\multicolumn{1}{|c}{69.1}&\multicolumn{1}{c}{65.5}&\multicolumn{1}{c}{\textbf{72.1}}&\multicolumn{1}{c}{68.6}\\ 
		
\multicolumn{1}{l|}{(\textit{Change})}&\multicolumn{1}{c}{(0)}&\multicolumn{1}{c}{($\downarrow$8.4)}&\multicolumn{1}{c}{($\uparrow$0.6)}&\multicolumn{1}{c}{($\downarrow$3.9)}&\multicolumn{1}{|c}{($\downarrow$1.9)}&\multicolumn{1}{c}{($\downarrow$10.1)}&\multicolumn{1}{c}{($\uparrow$0.6)}&\multicolumn{1}{c}{($\downarrow$4.9)}&\multicolumn{1}{|c}{($\downarrow$4.3)}&\multicolumn{1}{c}{($\downarrow$5.8)}&\multicolumn{1}{c}{($\uparrow$2.5)}&\multicolumn{1}{c}{($\downarrow$1.7)}\\ 
\hline


\end{tabular}
}
\vspace{-3mm}
\end{table*}


\subsubsection{Result} Table~\ref{tab:ablation} shows the result of the ablation study.  
Removing the DSL module(\textbf{w/o DSL}) results in a substantial decrease in accuracy (↓18.9--40.0\%) and F1-score (↓14.5--24.9\%), highlighting the crucial role of the structured, readable, and tool-agnostic DSL in guiding accurate configuration generation. 
The configuration name selector module also proves essential: its removal(\textbf{w/o Selector}) leads to severe drops in recall (↓74.2\% at the configuration name level and ↓62.7\% at the option value level), suggesting its importance in narrowing the generation space and improving coverage. 
While the alignment checker (\textbf{w/o Checker}) has a smaller overall impact, it still improves precision, with its absence causing up to a 10.1\% drop across all levels. 
The checker enhances the reliability of the configurations. 
The slight increase or stability in recall is expected, as the checker filters incorrect configurations rather than expanding coverage. 
Given that incorrect configurations can lead to misleading enforcement, precision is more critical than recall in this context.

\begin{tcolorbox}[width=5.5in,boxsep=1pt,left=1pt,right=1pt,top=1pt,bottom=1pt,colback  = white,colframe=myblue,, boxrule=0.3mm]
\textbf{Summary:} The ablation results for Checkstyle configuration generation in Java coding standards highlight the critical role of DSL-driven representations and AI-chain compilation.
\end{tcolorbox}
\vspace{-1mm}
\subsection{RQ4: Usefulness of Our Approach}\label{userstudy}
\subsubsection{Motivation} 
Given the challenges of manual linter configuration for coding standards in Section~\ref{motivation}, we evaluate whether our approach enables developers to generate these configurations more effectively. 
\subsubsection{Approach} 
We randomly select six various coding standards from the Google Java coding standard: two without corresponding Checkstyle configurations, two with a single configuration, and two with multiple configurations. 
Fourteen participants (seven PhD students and seven industry professionals, each with 1–5 years of Checkstyle experience) are recruited. 
Each participant is asked to write linter configurations for each coding standard in ten minutes. 
The participants are paired with similar backgrounds, one assigned to the control group (G1) and the other to the experimental group (G2). 
G1 is only given the coding standards, while G2 is also provided with configurations automatically generated by our approach as references. 
Participants are free to use the Internet or LLMs during the task, mirroring real‐world conditions. 
This setup allows us to evaluate whether our generated configurations can help developers effectively generate configurations in the wild. 
After the study, we conduct a short survey to collect feedback on participant behavior. 
We develop two separate online user study platforms for G1 and G2 to complete the task~\cite{g1_link, g2_link}.

To evaluate the performance difference between two groups, we compute their completion time and answer correctness. 
The completion time is automatically recorded during the study. 
Then we calculate the answer correctness with the percentage of questions answered correctly for each group (G1 and G2) at three levels: module name, option name, and option value. 
We use Wilcoxon signed-rank test~\cite{wilcoxon1992individual} to determine if the performance difference between two groups is statistically significant at the confidence level of 95\%.

\begin{table*}
\centering
\caption{Results of user study. 
} 
\vspace{-2mm}
\label{tab:userstudy_res}
\centering
\scalebox{0.76}{
\begin{tabular}{c|ccc|c}

\hline
\cellcolor{aliceblue}
&
\multicolumn{3}{c|}{\cellcolor{aliceblue}Correctness (\%)} &
 \cellcolor{aliceblue}
 \\
\cline{2-4}\multirow{-2}{*}{{\cellcolor{aliceblue}}Groups} &
\multicolumn{1}{c}{\cellcolor{aliceblue}Config Name Level} &
\multicolumn{1}{c}{\cellcolor{aliceblue}Option Name Level} &
\multicolumn{1}{c|}{\cellcolor{aliceblue}Option Value Level} 
&\multirow{-2}{*}{\cellcolor{aliceblue}Time(s)}
\\ \hline
G1&45.2&33.3&23.8&337.7 \\
\cline{1-5} 
G2&95.2&95.2&95.2&226.5\\
\cline{1-5} 
Change(\%)&$\uparrow$110.6&$\uparrow$185.9&$\uparrow$300.0&$\downarrow$49.1 
\\ \hline

\end{tabular}
}
\vspace{-4mm}
\end{table*}

\subsubsection{Result} 
Table~\ref{tab:userstudy_res} presents the results of the user study. 
\textit{Correctness} refers to the average percentage of correctly answered questions for each group across different configuration levels. 
\textit{Time} denotes the average completion time of a question for each group. 
The Wilcoxon signed-rank test shows statistically significant differences in correctness and completion time between G1 and G2 (p-value < 0.05), indicating that our approach helps developers configure linters more accurately and quickly. 

Participants in G1, who lacked reference configurations, exhibit declining correctness as the configuration granularity increased—from 45.2\% at the configuration name level to 23.8\% at the option value level. 
Interviews reveal that novice Checkstyle users struggled to identify which modules corresponded to which coding standards, citing the overwhelming number of available modules. 
Experienced users reported that identifying configuration names was relatively manageable, but accurately setting options was more challenging due to subtle details in the coding standards, especially under the 10-minute time constraint. 

G1 participants complete each task on average 337.7 seconds (five minutes and 37 seconds), which is significantly shorter than ten minutes.  
However, interviews reveal that this was not due to confidence or efficiency. 
Participants often ignore or misunderstand details in coding standards or linter configurations, such as checked objects of coding rules. 
Moreover, when they struggle to find relevant modules or precise option settings, they give up or submit answers to avoid running out of time. 

In contrast, group G2, equipped with our generated configuration references, achieves consistently high correctness across all levels (95.2\% at all granularities). 
Compared to G1, the minimum improvement in correctness exceeds 100\%. 
Moreover, G2 complete each task in an average of 226.5 seconds, 49.1\% faster than G1. 
All participants praise the references for reducing lookup time and improving configuration accuracy, highlighting their value in both correctness and efficiency. 
Across all 42 tasks completed by seven participants, only two are answered incorrectly, both by novices.
These errors arose when unsupported coding standards were erroneously assumed to be covered by Checkstyle. 
For example, for an ``empty block'' coding standard about brace formatting of empty blocks, a participant mistakenly generated the \textit{EmptyBlock} configuration.  
Although this configuration does not exist in our reference, the participant is misled by the keyword match in the module name and overlooks that \textit{EmptyBlock} actually prohibits empty statements in blocks, thus incorrectly considering it a valid configuration. 

\vspace{-1mm}
\begin{tcolorbox}[width=5.5in,boxsep=1pt,left=1pt,right=1pt,top=1pt,bottom=1pt,colframe = myblue, colback  = white,boxrule=0.3mm]
\textbf{Summary:} The Checkstyle configurations generated by our approach effectively help developers configure linters more accurately and efficiently. 
\end{tcolorbox}

\subsection{RQ5: Generality of Our Approach}\label{usefulness}

\subsubsection{Motivation} We are interested in whether our approach can be effectively extended to other programming languages, coding standards and linters.

\subsubsection{Approach} \label{rq3_aaproach}
To robustly assess generality while keeping benchmarking effort manageable, we prioritize selecting different programming languages, linters, and coding standards based on the complexity of the task. 
We choose JavaScript, a dynamically-typed, interpreted language primarily used for web development, contrasting with Java, a strongly-typed, compiled language used in enterprise and mobile applications. 
For linters, we select ESLint, which uses JSON, in contrast to Checkstyle's XML-based configurations. 
We use the Google JavaScript style guide as the Javascript coding standard for our study. 
The Google JavaScript style guide covers a broader set of rules, with twice as many coding standards as the Java coding standard, providing a richer dataset to evaluate the generality of our approach. 
Additionally, although the JavaScript style guide is also from Google, their natural language descriptions differ due to the distinct programming languages. 
Even though for similar coding rules, their description is different. 
For instance, the Java standard states, ``The column limit (Section 4.4, Column limit: 100) does not apply to import statements'', while JavaScript's standard says, ``Import statements must not be line-wrapped and are therefore an exception to the 80-column limit''. 
When applying the approach to other programming languages, coding standards, and linter configurations, we only need to provide the coding standards, linter documentation, and the linter configuration format (e.g., JSON) to directly utilize our approach. 
The evaluation approach, dataset, baselines, and metrics are consistent with those described in Sections~\ref{rq_dsl_approach} and~\ref{rq1_approa}.

Table~\ref{tab:benchmarkl_js_eslint} shows the benchmark of ESLint configurations for JavaScript coding standards, which took about 182 hours to create. 
For the Google JavaScript style guide, 59.7\% of the 149 coding standards lacked configurations, 40.3\% had configurations,  of which 30.2\% required options or multiple configuration names. 
The benchmark also highlight the challenge in generating linter configurations, as many coding standards either lack corresponding configurations or require complex configurations with options or with multiple configuration names.

\begin{table}[t]
\caption{Benchmark statistics for ESLint configurations on Google JavaScript coding standard.} 
\vspace{-2mm}
\label{tab:benchmarkl_js_eslint}
\centering
\scalebox{0.8}{
\begin{tabular}{ll}
\multicolumn{1}{c|}{\cellcolor{aliceblue} Configuration Category for Coding Standard }
&\multicolumn{1}{c}{\cellcolor{aliceblue}Number}\\ 
\hline
\multicolumn{1}{l|}{No Configuration} 
&\multicolumn{1}{c}{89} \\\hline

\multicolumn{1}{l|}{Has Configuration} 
&\multicolumn{1}{c}{60} \\\hline

\multicolumn{1}{l|}{Config with Only Configuration Name} 
&\multicolumn{1}{c}{15}\\\hline

\multicolumn{1}{l|}{Config with Configuration Name and Options} 
&\multicolumn{1}{c}{45}\\\hline

\multicolumn{1}{l|}{Config with Multiple Configurations} 
&\multicolumn{1}{c}{16}\\\hline

\multicolumn{1}{l|}{Total Number of Coding Standards} 
&\multicolumn{1}{c}{149}\\
\hline

\end{tabular}
}
\vspace{-3mm}
\end{table}

\begin{table}
\caption{Results of DSL generation for JavaScript coding standards and ESLint configurations. 
}
\vspace{-0.2cm}
\label{tab:metric_dsl_js_eslint}
\centering
\scalebox{0.8}{
\begin{tabular}{llllllll}
\multicolumn{1}{c|}{\cellcolor{aliceblue}Name}&
\multicolumn{1}{c}{\cellcolor{aliceblue}\#N}&
\multicolumn{1}{c}{\cellcolor{aliceblue}\#S}&
\multicolumn{1}{c}{\cellcolor{aliceblue}\#DSL}&
\multicolumn{1}{c}{\cellcolor{aliceblue}Acc(\%)}&
\multicolumn{1}{c}{\cellcolor{aliceblue}P(\%)}&
\multicolumn{1}{c}{\cellcolor{aliceblue}R(\%)}&
\multicolumn{1}{c}{\cellcolor{aliceblue}F1(\%)}\\ \hline

\multicolumn{1}{c|}{JS Style Guide}&
\multicolumn{1}{c}{149}&
\multicolumn{1}{c}{108}&
\multicolumn{1}{c}{393}&
\multicolumn{1}{c}{84.3}&
\multicolumn{1}{c}{94.6}&
\multicolumn{1}{c}{99.5}&
\multicolumn{1}{c}{97.0}\\ 
\hline

\multicolumn{1}{c|}{ESLint}&
\multicolumn{1}{c}{291}&
\multicolumn{1}{c}{166}&
\multicolumn{1}{c}{540}&
\multicolumn{1}{c}{84.9}&
\multicolumn{1}{c}{92.7}&
\multicolumn{1}{c}{99.4}&
\multicolumn{1}{c}{96.0}\\ 
\hline

\end{tabular}
}
\end{table}

\begin{table*}
\centering
\caption{Results of generating ESLint configurations for Google JavaScript coding standards.}
\vspace{-2mm}
\label{tab:metrics_resl_js_eslint}
\centering
\scalebox{0.7}{
\begin{tabular}{c|llllllllllll|}

\hline
\cellcolor{aliceblue} &
\multicolumn{4}{c}{\cellcolor{aliceblue}Config Name Level Metrics(\%)} &
\multicolumn{4}{|c}{\cellcolor{aliceblue}Option Name Level Metrics(\%)} &
\multicolumn{4}{|c}{\cellcolor{aliceblue}Option Value Level Metrics(\%)} \\
\cline{2-13} \multirow{-2}{*}{\cellcolor{aliceblue}Approach} &
\multicolumn{1}{c}{\cellcolor{aliceblue}Acc} &
\multicolumn{1}{c}{\cellcolor{aliceblue}P} &
\multicolumn{1}{c}{\cellcolor{aliceblue}R} &
\multicolumn{1}{c}{\cellcolor{aliceblue}F1} &
\multicolumn{1}{|c}{\cellcolor{aliceblue}Acc} &
\multicolumn{1}{c}{\cellcolor{aliceblue}P} &
\multicolumn{1}{c}{\cellcolor{aliceblue}R} &
\multicolumn{1}{c}{\cellcolor{aliceblue}F1} &
\multicolumn{1}{|c}{\cellcolor{aliceblue}Acc} &
\multicolumn{1}{c}{\cellcolor{aliceblue}P} &
\multicolumn{1}{c}{\cellcolor{aliceblue}R} &
\multicolumn{1}{c}{\cellcolor{aliceblue}F1} \\ \hline

\multicolumn{1}{l|}{Closed-book}&\multicolumn{1}{c}{28.9}&\multicolumn{1}{c}{21.3}&\multicolumn{1}{c}{68.5}&\multicolumn{1}{c}{32.5}&\multicolumn{1}{|c}{24.8}&\multicolumn{1}{c}{14.1}&\multicolumn{1}{c}{45.0}&\multicolumn{1}{c}{21.4}&\multicolumn{1}{|c}{24.8}&\multicolumn{1}{c}{12.4}&\multicolumn{1}{c}{39.8}&\multicolumn{1}{c}{18.9}\\ \hline

\multicolumn{1}{l|}{Name}&\multicolumn{1}{c}{20.8}&\multicolumn{1}{c}{12.8}&\multicolumn{1}{c}{77.8}&\multicolumn{1}{c}{22.0}&\multicolumn{1}{|c}{16.8}&\multicolumn{1}{c}{8.5}&\multicolumn{1}{c}{51.9}&\multicolumn{1}{c}{14.6}&\multicolumn{1}{|c}{16.1}&\multicolumn{1}{c}{7.8}&\multicolumn{1}{c}{47.2}&\multicolumn{1}{c}{13.3}\\ \hline

\multicolumn{1}{l|}{Name+Desc}&\multicolumn{1}{c}{\underline{36.2}}&\multicolumn{1}{c}{\underline{24.3}}&\multicolumn{1}{c}{64.8}&\multicolumn{1}{c}{\underline{35.4}}&\multicolumn{1}{|c}{\underline{28.9}}&\multicolumn{1}{c}{\underline{16.7}}&\multicolumn{1}{c}{44.4}&\multicolumn{1}{c}{\underline{24.2}}&\multicolumn{1}{|c}{\underline{28.2}}&\multicolumn{1}{c}{\underline{15.6}}&\multicolumn{1}{c}{41.7}&\multicolumn{1}{c}{\underline{22.7}}\\ \hline

\multicolumn{1}{l|}{Name+Desc+Opts}&\multicolumn{1}{c}{-}&\multicolumn{1}{c}{-}&\multicolumn{1}{c}{-}&\multicolumn{1}{c}{-}&\multicolumn{1}{|c}{-}&\multicolumn{1}{c}{-}&\multicolumn{1}{c}{-}&\multicolumn{1}{c}{-}&\multicolumn{1}{|c}{-}&\multicolumn{1}{c}{-}&\multicolumn{1}{c}{-}&\multicolumn{1}{c}{-}\\ \hline

\multicolumn{1}{l|}{RAG (Name+Desc)}&\multicolumn{1}{c}{14.8}&\multicolumn{1}{c}{11.7}&\multicolumn{1}{c}{77.8}&\multicolumn{1}{c}{20.4}&\multicolumn{1}{|c}{10.7}&\multicolumn{1}{c}{7.4}&\multicolumn{1}{c}{49.1}&\multicolumn{1}{c}{12.8}&\multicolumn{1}{|c}{10.7}&\multicolumn{1}{c}{6.8}&\multicolumn{1}{c}{45.4}&\multicolumn{1}{c}{11.9}\\ \hline

\multicolumn{1}{l|}{RAG (Name+Desc+Opts)}&\multicolumn{1}{c}{18.1}&\multicolumn{1}{c}{9.6}&\multicolumn{1}{c}{\textbf{\underline{80.6}}}&\multicolumn{1}{c}{17.2}&\multicolumn{1}{|c}{14.8}&\multicolumn{1}{c}{6.6}&\multicolumn{1}{c}{\underline{55.6}}&\multicolumn{1}{c}{11.8}&\multicolumn{1}{|c}{14.1}&\multicolumn{1}{c}{6.2}&\multicolumn{1}{c}{\underline{51.9}}&\multicolumn{1}{c}{11.1}\\ \hline

\multicolumn{1}{l|}{\textbf{Our Method}}&\multicolumn{1}{c}{\textbf{75.2}}&\multicolumn{1}{c}{\textbf{84.4}}&\multicolumn{1}{c}{70.4}&\multicolumn{1}{c}{\textbf{76.8}}&\multicolumn{1}{|c}{\textbf{71.1}}&\multicolumn{1}{c}{\textbf{75.6}}&\multicolumn{1}{c}{\textbf{63.0}}&\multicolumn{1}{c}{\textbf{68.7}}&\multicolumn{1}{|c}{\textbf{71.1}}&\multicolumn{1}{c}{\textbf{75.6}}&\multicolumn{1}{c}{\textbf{63.0}}&\multicolumn{1}{c}{\textbf{68.7}}\\ \hline

\multicolumn{1}{l|}{Change (\%)}&\multicolumn{1}{c}{$\uparrow$107.7}&\multicolumn{1}{c}{$\uparrow$247.3}&\multicolumn{1}{c}{$\downarrow$12.7}&\multicolumn{1}{c}{$\uparrow$116.9}&\multicolumn{1}{|c}{$\uparrow$146.0}&\multicolumn{1}{c}{$\uparrow$352.7}&\multicolumn{1}{c}{$\uparrow$13.3}&\multicolumn{1}{c}{$\uparrow$183.9}&\multicolumn{1}{|c}{$\uparrow$152.1}&\multicolumn{1}{c}{$\uparrow$384.6}&\multicolumn{1}{c}{$\uparrow$21.4}&\multicolumn{1}{c}{$\uparrow$202.6}\\ \hline

\multicolumn{13}{p{400pt}}{Note:  `-' means inapplicable results due to LLMs' token limits. }\\

\end{tabular}
}
\vspace{-6mm}
\end{table*}

\subsubsection{Result} 
Table~\ref{tab:metric_dsl_js_eslint} presents the accuracy, precision, recall, and F1-score for generating DSL representations of JavaScript coding standards and ESLint configurations. 
We review 393 DSL representations for 108 coding standards and 540 for 166 tool configurations. Precision, recall, and F1 scores consistently exceed 90\%, with accuracy above 80\%. 
The results highlight the generality of our approach in generating DSL representations. 
 
For ESLint configurations of JavaScript coding standards, our approach achieves 71.1\%$\sim$75.2\% for accuracy, 75.6\%$\sim$84.4\% for precision, 63.0\%$\sim$70.4\% for recall, and 68.7\%$\sim$76.8\% for F1-score across three levels of granularity. 
Similar to the results for Checkstyle configurations of Java coding standards, our approach consistently outperforms the baselines in accuracy, precision, and F1-score across all granularity levels, with minimum improvements of 107.7\%, 247.3\% and 116.9\%. 
While recall decreases at the configuration name level (-12.7\%), it improves at the option name and option value levels (+13.3\% and +21.4\%, respectively). 
Notably, precision is more critical than recall in tool configurations, ensuring correctness and minimizing deployment errors. 
As with Checkstyle configurations, our approach significantly improves precision (by more than 100\%) while maintaining strong recall.

\begin{tcolorbox}[width=5.5in,boxsep=1pt,left=1pt,right=1pt,top=1pt,bottom=1pt,colframe = myblue, colback  = white,boxrule=0.3mm]
\textbf{Summary:} The high accuracy, F1-score, precision and recall of ESLint configurations for JavaScript coding standards illustrate that our approach can be effectively extended to other programming languages, coding standards and linters. 
\end{tcolorbox}

\section{Discussion}\label{discussion}
\subsection{Implications} 

Our approach, detailed in Section~\ref{method} and evaluated in Section~\ref{ourevaluation}, demonstrates promising performance and scalability potential in linter configuration generation for coding standards. 
\textbf{It can be readily adopted by software companies, large-scale projects, and individual developers to generate and maintain configurations for coding standards}. 
 To apply it to other programming languages, coding standards, or linters, developers only need to provide the coding standards, linter documentation, and configuration format (e.g., XML). 
%

The design core is the DSL-driven compilation pipeline with LLMs. 
Inspired its effectiveness, a \textbf{new insight for solving domain-related tasks} is that, instead of leaving everything to the LLM or fine-tuning LLMs with datasets, it is crucial to \textbf{design a lightweight DSL that models knowledge and information} in a structured, precise, and readable way. 
Since natural language is free-form, unstructured, and often ambiguous, designing the DSL can enhance the clarity and precision of communication with LLMs. 
By building on LLMs' strengths in generation, we can have LLMs function more as a subroutine instead of the central driver, ultimately improving the ability to address domain-specific tasks.

Meanwhile, our compilation pipeline leverages core compiler concepts—such as syntax parsing, intermediate code generation, semantic analysis, and machine code generation—to ensure modularity, reusability, and extensibility. 
For example, if coding standards have incomplete information, we could add a module for information completion before the NL coding standard parser. 
If we want to enhance semantic reasoning capabilities, we could introduce a reasoning module before or during the DSL instruction configurator. 
Besides, unlike traditional compilers, we leverage LLMs to enable fine-grained linter configuration. 
\textbf{Future research can explore the potential of LLMs as compilers} independent of programming languages, operating systems, and hardwares.

Finally, the approach and benchmark can assist \textbf{future research in conducting comparative studies} on the capabilities of various linters and their alignment with coding standards.
They can also \textbf{facilitate the collaborative development of linters and coding standards}.  
For example, our benchmarks uncover limitations in existing linters: 27.9\% of coding standards from the Google Java coding standard and 59.7\% of coding standards from the Google JavaScript coding standard are unsupported by Checkstyle and ESLint in Section~\ref{rq1_approa}, Section~\ref{rq3_aaproach}, Table~\ref{tab:benchmark} and Table~\ref{tab:benchmarkl_js_eslint}. 
We further analyze the unsupported coding standards in linters and identify two reasons. 
First, coding standards may require deep code analysis beyond current linters' capabilities. 
For example, a Java coding standard that static imports must not reference static nested classes requires cross-file resolution, which Checkstyle cannot perform. 
Second, a single conceptual coding standard may have multiple syntactic variants, but linters typically support only some of them. 
For instance, the JavaScript coding standard specifies indentation for continuation lines, yet ESLint provides no configuration for it. 
These unsupported standards can be reasonably supported, but doing so would demand significant implementation effort and extensions to linter configuration designs. 
Researchers and linter developers can use these findings to prioritize improvements and extend linter capabilities for better alignment with coding standards.

\vspace{-2mm}
\subsection{Threats to Validity}
\noindent \textbf{Internal Validity:}
\noindent A possible threat to internal validity is the presence of human errors in the manual evaluation in Section~\ref{ourevaluation}. 
To reduce the impact of the problem, each instance was evaluated by at least two authors and two external experts. 
In case of disagreement, all the evaluators revisited the instances to reach a consensus. 

\noindent \textbf{External Validity:} 
One potential threat to external validity lies in our selection of programming languages, coding standards, and linter configurations. 
However, our goal is to evaluate the feasibility and generalizability of the proposed method, not to achieve exhaustive coverage. 
Our experiments validate its effectiveness across Java and JavaScript, the Google Java and JavaScript coding standards with different coding standards and description, and Checkstyle and ESLint with different linter formats. 
This design enables us to demonstrate that our approach is not tied to a specific programming language, coding standards, or linters. 
Benchmark construction is another challenge, requiring substantial effort—we spent 302 hours manually creating these benchmarks. 
In the future, our approach can help construct additional benchmarks, enabling broader coverage of programming languages, coding standards, and linters. 
\section{Related Work}\label{relatedwork}


\noindent \textbf{Studies on Coding Standards.} 
Coding standards evolve as organizations refine best practices, adopt new paradigms, and phase out outdated rules~\cite{pruim2023fostering,butler2015investigating,abdallah2017java,simmons2020large,boogerd2008assessing,smit2011code,dos2018impacts,dos2018impacts,oliveira2022lint}. Abdallah et al.~\cite{abdallah2017java} analyzed 20 widely-used Java coding standards from 1996 to 2016. 
Their study revealed that organizations often develop their own standards, which evolve over time through the addition of new coding standards, the removal of outdated ones, and the updating to existing coding standards. 
Among these, the SUN and Google standards were particularly effective. 
Butler et al.~\cite{butler2015investigating} investigated Java naming conventions and found that while developers generally adhere to naming conventions, compliance with specific guidelines varies significantly. 
Beyond empirical studies, researchers have explored mining coding standards~\cite{hora2013mining,campos2019mining,latappy2023mlinter,allamanis2014learning,markovtsev2019style,smit2011code}. 
Allamanis et al.~\cite{allamanis2014learning} introduced NATURALIZE, leveraging statistical NLP to infer coding style consistency. 
Unlike prior work, we propose a DSL to express coding rules  independently of programming languages, coding standards and linters.

\noindent \textbf{Studies on Linters for Coding Standards.} 
Numerous linter tools have been developed to improve code quality and reduce developer effort~\cite{ogura2018bring,hart2023eastwood,allamanis2014learning,yiu2023checkstyle,dilruk2019coding,livshits2005finding,christodorescu2003static,tsantalis2008jdeodorant,latappy2023mlinter,oumarou2015identifying,pmd}. 
Well-known researched static analysis tools include FindBugs~\cite{findbugs}, Checkstyle~\cite{ckstyle} and ESLint~\cite{eslint}. 
Beyond tool development, studies have examined developer usage and challenges in adopting these tools~\cite{beller2016analyzing,zampetti2017open,christakis2016developers,tomasdottir2020adoption,tomasdottir2017and}. 
Beller et al.~\cite{beller2016analyzing} found that developers frequently modify default configurations and prioritize minimizing false positives. 
Tómasdóttir et al.~\cite{tomasdottir2020adoption} analyzed over 9,500 ESLint configurations and surveyed developers, revealing that evolving coding standards complicate tool adoption due to inconsistencies in rule interpretation and integration challenges. 
Unlike prior work, which focuses on tool usage and static analysis, we generate  tool configurations to relieve developers from the burden of manual tool configuration.

\noindent \textbf{Studies on DSL.} 
Domain-specific languages (DSLs) offer significant benefits by providing a high-level abstraction for specifying problems within particular domains~\cite{gray2008dsls, bragancca2021towards, ries2018messir, gandhi2023natural, schafer2011pattern, desai2016program}. 
Historically, translating natural language into DSLs has been complex. 
Desai et al.~\cite{desai2016program} developed a synthesis algorithm to translate English sentences into DSL programs, incorporating a training phase to learn a dictionary and weights for the algorithm. 
With the advent of LLMs, Gandhi et al.~\cite{gandhi2023natural} demonstrated how LLMs can translate natural language inputs into DSL programs that interact with application APIs. 
Unlike these approaches, 
we start by designing a tool-independent DSL structurally and precisely represent coding rules from coding standards and linter configurations, followed by utilizing an AI-based compilation pipeline to generate linter configurations for coding standards.

\vspace{-1.3mm}

\section{Conclusion}\label{conclusion}
We design a domain-specific language (DSL) and employ an AI chain compilation to automate the generation of linter configurations for coding standards. 
The DSL express coding rules of coding standards and linter configurations in a tool-independent, structured, precise and readable way. 
 The AI chain transpilation ensures the generation of accurate and fine-grained linter configurations. 
Experiments show the effectiveness, usefulness and generality of the designed DSL and our approach. 
In the future, we will employ our approach to assist constructing more benchmarks with more programming languages, coding standards and linters, followed by empirical studies to align coding standards with linters, fostering collaborative development of coding standards and linters. 

\section{Data Availability}\label{data}
The source code and data can be found here~\cite{Replication_Package}. 
\bibliographystyle{ACM-Reference-Format}

\bibliography{sample-base}




\end{document}